\documentclass[acmlarge]{acmart}
\makeatletter
\newcommand{\myconfshort}{\acmConference@shortname}
\newcommand{\myconffull}{\acmConference@name}
\newcommand{\myconfdate}{\acmConference@date}
\newcommand{\myconfloc}{\acmConference@venue}
\AtBeginDocument{
  \fancypagestyle{firstpagestyle}{
    \fancyhead{}%
    \fancyfoot[C]{}%
  }
  \fancyhf{}
  \fancyhead[LO]{\@headfootfont\shorttitle}%
  \fancyhead[RE]{\@headfootfont\@shortauthors}%
  \fancyhead[LE]{\@headfootfont\footnotesize \myconfshort, \myconfdate, \myconfloc}%
  \fancyhead[RO]{\@headfootfont\footnotesize \myconfshort, \myconfdate, \myconfloc}%
  \fancyfoot[C]{}%
}
\makeatother
\usepackage{tabularx}
\usepackage{verbatim}
\usepackage{comment}
\usepackage{caption} 
\AtBeginDocument{%
  }

\copyrightyear{2026}
\acmYear{2026}
\setcopyright{none}
\acmConference[FAccT '26]{The 2026 ACM Conference on Fairness, Accountability, and Transparency}{June 25--28, 2026}{Montreal, QC, Canada}
\acmBooktitle{The 2026 ACM Conference on Fairness, Accountability, and Transparency (FAccT '26), June 25--28, 2026, Montreal, QC, Canada}
\acmDOI{10.1145/3805689.3812258}
\acmISBN{979-8-4007-2596-8/2026/06}

\renewcommand\footnotetextcopyrightpermission[1]{}
\begin{document}

\title{How Formerly Incarcerated People Envision Technologies for\\ Prison Parole}

\author{Saiph Savage}
\email{s.savage@northeastern.edu}
\affiliation{%
  \institution{Northeastern University}
  \city{Boston}
  \state{Massachusetts}
  \country{USA}
}

\author{Jesse Nava}
\affiliation{%
  \institution{San Diego State University (SDSU)}
  \city{San Diego}
  \state{California}
  \country{USA}
}


\author{Wanqing Iris Zhou}
\affiliation{
 \institution{Brandeis University}
 \city{Boston}
 \state{Massachusetts}
 \country{USA}}

\author{HwiJoon Lee}
\email{lee.hw@northeastern.edu}
\affiliation{%
  \institution{Northeastern University}
  \city{Boston}
  \state{Massachusetts}
  \country{USA}
  }

\newcommand{\rev}[1]{\textcolor{black}{#1}}
\renewcommand{\shortauthors}{Savage et al.}

\begin{teaserfigure}
  \centering
  \includegraphics[width=0.9\textwidth]{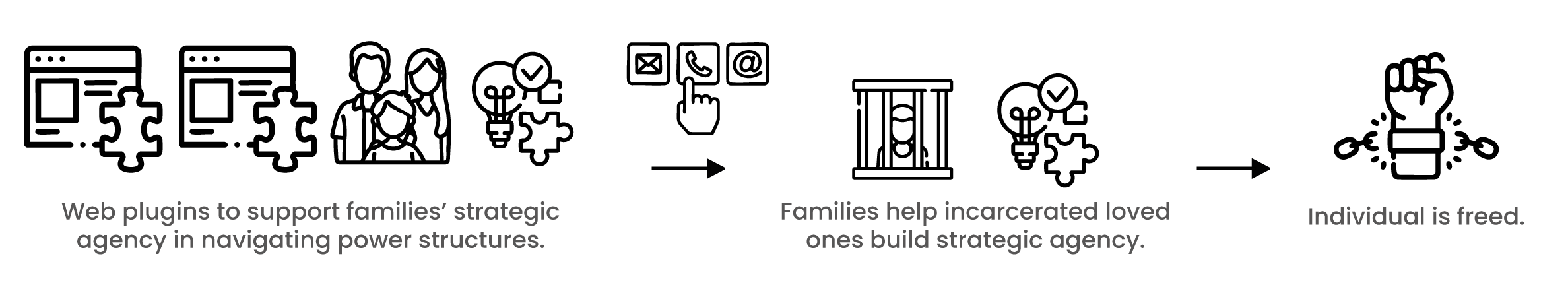}
  \caption{Overview of our study's proposed design direction for future computational tools to support parole.}
  \Description{A sequence diagram illustrating a process of empowerment and release: On the left, web plugins are shown supporting families’ strategic agency in navigating power structures (depicted by browser windows, puzzle pieces, and a family icon). Arrows lead to communication tools (email, phone, messaging), then to an image of a jailed individual, indicating families helping incarcerated loved ones build strategic agency. The final stage shows a person breaking free from handcuffs, representing the individual gaining freedom.}
  \label{fig:teaser}
\end{teaserfigure}

\begin{abstract}
AI-driven algorithms and automated tools are increasingly embedded in the correctional landscape, shaping parole eligibility, release decisions, and surveillance. These tools are also often framed as objective, inevitable solutions to inefficiency and bias. Yet, these computational systems are rarely designed with input from justice-impacted individuals, which means they might fail to address the real needs of incarcerated people. To address this gap, we surveyed 31 formerly incarcerated people about their parole experiences and their visions for technologies that could support parole preparation. Contrary to dominant assumptions, participants did not imagine computational tools as instruments to dismantle the prison system, but as resources for navigating power: translating complex parole concepts into culturally familiar terms, documenting personal transformation in board-legible ways, and recognizing the often-invisible labor of families. We argue that these imaginaries point toward technologies designed for strategic agency, where tools help incarcerated individuals and their families build the capacity to navigate existing power structures in pursuit of freedom. We conclude by reframing computational tools for parole away from surveillance and toward human-centered systems that support people in navigating carceral power structures.
\end{abstract}



\maketitle

\section{Introduction}
Parole is the conditional release of an individual from prison under community supervision \cite{phelps2017supervision,solomon2005does}, contingent on compliance with rules set by a parole board \cite{bjs_glossary,uspc_faq_2025}. Over the past decade, automated and AI-driven systems have started to become deeply woven into the governance of parole \cite{stevenson2024algorithmic,solow2019institutional}. 
\rev{This trend has also broadly extended to the integration of computational systems into legal decision-making \cite{assamidanov2023striking}. Correctional agencies across the United States now routinely depend on computational tools to estimate recidivism risk, guide release decisions, determine supervision intensity, allocate rehabilitative resources, and manage caseloads \cite{berk2017impact,dressel2018accuracy,slobogin2021preventive}. Beyond the release decision itself, these computational system are increasingly shaping the everyday conditions of supervision through electronic monitoring, automated reporting, and digital compliance workflows that structure how people on parole are evaluated, questioned, and sanctioned \cite{viglione2024expansion,al2025effects}.}
For example, judges across the United States are beginning to use AI to summarize filings, review evidence, and map out draft opinions \cite{cheref2024delicate,remus2017can,sourdin2018judge}. Entire state court systems are piloting AI tools, and federal judges now have access to generative AI platforms designed specifically for judicial decision support \cite{reiling2020courts,nowotko2021ai}. Even arbitration bodies have begun offering AI-generated decisions in certain types of disputes \cite{solhchi2023artificial,katsh2017digital}. This growing normalization of computational tools within the legal system has been reinforcing a narrative in which algorithmic systems are framed as both “objective” and “inevitable” solutions to long-standing problems of backlog, inconsistency, and human error \cite{Hlavca19,wang20,hildebrandt2018algorithmic}. Similar narratives have started to surround automated risk assessment tools that are used in parole, which are also frequently promoted as more ``accurate'' and ``neutral'' than human judgment \cite{dressel2018accuracy,jones2007probation,stevenson2024algorithmic,paparozzi2008probation}. 

Despite expanding reliance on computational tools in parole and enthusiasm about its potential \cite{slobogin2021preventive,scaria2024algorithms,talukderartificial,nellis2022electronic}, justice-impacted individuals are almost entirely excluded from the design, evaluation, and oversight of these systems \cite{schwerzmann2021abolish}. This exclusion has consequences. Without the input of people who are directly affected, computational systems can encode institutional goals and assumptions that do not reflect the lived realities of the people actually navigating parole \cite{eubanks2018automating,brayne2021predict,benjamin2023race}. For example, many AI-enhanced risk assessment tools used in parole decisions rely on variables such as “housing stability,” “prior employment,” “family circumstances,” and “neighborhood environment” \cite{joseph2025predicting,dressel2018accuracy}. Although these variables are presented as neutral predictors of who is more likely to violate parole conditions, they are also well documented to encode racial and socioeconomic inequities \cite{angwin2022machine,skeem2016risk}. As a result, the computational tools used in the parole process can reproduce historical disparities rather than correct them \cite{Meyer22,Zilka23,zhang24}. The effects are not abstract. For example, a risk assessment model could mistakenly label a parolee's proposed housing as “high risk” simply because the home is in a low-income neighborhood, even when that housing is stable and supportive for them. This misclassification can shrink the person's viable release options, sever access to support networks, and treat culturally grounded forms of care as illegible or unacceptable to the system \cite{herbert2015homelessness,bryan2023housing,phelps2013paradox}. 

When justice-impacted voices are absent from system design, parole technologies usually tend to prioritize institutional goals such as surveillance, compliance, and administrative efficiency \cite{brayne2021predict,bagaric2018introducing,pattavina2004emerging}. As a result, these new computational systems can expand state control rather than support reintegration, creating a persistent mismatch between the needs of people on parole and the technological infrastructures that shape their access to freedom \cite{harding2022supervision}. In this context, these computational systems are not neutral \cite{benjamin2023race,pollicino2021getting,crawford2021atlas}. They become powerful actors that influence who is granted freedom, under what conditions, and according to whose values \cite{brayne2021predict,he2025algorithm,he2025algorithm}. For many justice-impacted individuals, computational tools are increasingly a recurring presence in their lives \cite{pattavina2004emerging,magassa2024inclusive}, but it is a presence that seldom operates in their interest or their families’ and instead tends to reinforce institutional priorities \cite{benjamin2023race,brayne2021predict}. \rev{This institutional orientation is also reflected in how prior scholarship has framed the role of technology in carceral settings.}

\rev{Prior criminal justice, HCI, and FAccT scholarship has often understood these technologies through three dominant lenses \cite{ramesh2022platform,devrio2024building,wieringa2020account,10.1145/3531146.3533194}. First, technologies are framed as tools for prediction and risk management \cite{zilka2023progression,berk2019machine,angwin2022bias}, such as risk assessment systems that estimate recidivism and inform release decisions \cite{angwin2022machine,berk2019machine}. Second, they are understood as mechanisms of surveillance and compliance \cite{brayne2021predict,ziosi2024evidence}, including electronic monitoring and digital reporting systems that extend institutional control beyond prison walls \cite{brayne2021predict}. Third, a growing body of FAccT scholarship critiques these systems through the lens of fairness, accountability, and transparency, focusing on bias, auditability, and the harms of algorithmic decision-making \cite{ehsan2022algorithmic,selbst2019fairness,raji2020closing}. Together, these approaches position technology primarily as either a tool of institutional control or a target of external critique. Although this work has been essential for showing how computational systems reproduce inequality \cite{benjamin2023race,Zilka23}, it leaves comparatively underexplored how justice-impacted individuals themselves imagine and engage with technology as a resource for navigating these systems in practice.}

This paper responds to this problem space by reimagining what technologies for parole might look like if the people most affected were given the opportunity to define technology’s role in the parole process. To do so, we surveyed 31 formerly incarcerated individuals about their experiences with parole and their visions for computational tools that could support parole preparation. We found that participants imagined tools that would help them better navigate power structures within the parole process. For example, they described how parole board commissioners often hold cultural stereotypes about how a “rehabilitated” person should speak, behave, and present their education. When participants did not match these expectations, they were perceived as ``manipulative'' and were consequently denied parole. But, when they did match stereotypical expectations, they were seen as lacking rehabilitation and as still deserving incarceration. In response to these biases, participants imagined technologies that could help them systematically document and quantify the pro-social activities they completed while in prison, so that their transformation would be visible and credible to decision-makers. Participants further described struggling with concepts that parole commissioners expected them to engage with, such as particular psychological or rehabilitative constructs that were culturally distant. They imagined tools that would translate these concepts into culturally accessible terms, helping them understand what the board valued and how to prepare more effectively for their parole board hearing. Finally, participants also emphasized technologies that could reduce the often-invisible labor that families and close supporters perform to help them comply with parole conditions, underscoring that navigating institutional demands is frequently a collective effort. We argue that the tools participants envisioned are tools for building ``strategic agency'' \cite{mahmood2016feminist}, understood as the capacity to pursue freedom by interpreting, anticipating, and effectively responding to institutional power and its expectations. Together, our findings show that participants’ imaginaries of technology focus less on escaping power and more on navigating it, through tools that make institutional rules more intelligible, progress more legible, and support networks more sustainable. Accordingly, we call for a reorientation of carceral technology design. Rather than concentrating on intensifying surveillance or prediction \cite{brayne2021predict,benjamin2023race}, we argue for computational tools that strengthen strategic agency for incarcerated individuals and their families, enabling them to navigate power structures more effectively. 

Figure \ref{fig:teaser} summarizes our study-informed design recommendations for parole-related technologies. We argue that these technologies should strengthen family members’ strategic agency to navigate carceral power structures on behalf of their incarcerated loved ones. We emphasize family-centered tools because technologies designed for direct use by incarcerated people often require prison approval \cite{metcalf2021algorithmic,manning2008technology}, which can slow adoption or create risks that carceral institutions reject or weaponize the tools \cite{payne1998qualitative,mcneill2018pervasive}. Based on our findings, we also argue that these tools should fit within the information ecosystems families already use, potentially taking the form of web browser plugins \cite{picazo2020after,cao2020activist,firmenich2022engineering}, rather than requiring families to adopt entirely new platforms. Finally, these tools should also guide family members in helping their incarcerated loved ones develop the strategic agency needed to pursue parole release. However, this guidance should be grounded in official prison communication channels so that families’ strategies are legitimate, actionable, and responsive to the constraints of the carceral system.
\newpage
In this paper, we contribute:
\begin{itemize}
\item {\bf Empirical contribution:} The experiences of formerly incarcerated individuals with the parole process and their imaginaries of technologies that could support the process.
\item {\bf Design contribution:} Reframing computational tools for parole away from risk assessment and surveillance, and toward supporting the capacity of justice-impacted individuals to navigate power structures.
\end{itemize}

\section{Related Work}
We situate our work within three strands of prior research: (1) the growing use of AI and computational tools in criminal justice and critical perspectives on its deployment; (2) technologies designed for incarcerated populations; and (3) the parole system as an understudied site for human-centered design.

\subsection{AI and Computational Tools in Criminal Justice}
AI-based technologies and computational tools, in general, are increasingly deployed across criminal justice contexts, from risk assessment instruments to predictive policing systems \cite{monahan2016risk, klingele2015promises}. Proponents argue that algorithmic tools can improve efficiency and consistency in decision-making \cite{monahan2016risk}, and some empirical work suggests algorithms may outperform human judgment in identifying low-risk defendants \cite{kleinberg2018human}. However, critical scholars have documented how these purported benefits are unevenly distributed. Risk assessment tools have been shown to perpetuate racial disparities despite limited predictive accuracy \cite{chouldechova2017fair, dressel2018accuracy}. Predictive policing systems reinforce over-policing in already marginalized communities \cite{lum2016predict}. Rather than reducing bias, scholars argue that technology in criminal justice has largely served to automate and obscure institutional power while generating a black box of objectivity \cite{eubanks2018automating}. These systems seem to focus on just extending surveillance, constraining individual agency, and further complicating carceral logics.

\subsection{Current Technology and Incarcerated Populations}
Incarcerated people are largely deprived of access to the internet and other information and communication technologies (ICTs) \cite{jewkes2016brave}. This creates a form of digital exclusion that can be especially severe in today’s digital age \cite{jewkes2016brave, reisdorf2016b, zivanai2022digital}. Some prisons have introduced self-service kiosks and tablets that provide educational materials, communication tools, and account management \cite{mcdougall2019technology, reisdorf2024locked, yildirim2024smart, mahlangu2023offender}. Advocates argue that ICT access can support rehabilitation by helping incarcerated individuals maintain family relationships and build stability, which can reduce recidivism \cite{sobol2018connecting}. The COVID-19 pandemic further demonstrated the feasibility and benefits of expanding digital access in prisons \cite{reisdorf2024locked}. However, these initiatives are often shaped by corporate monopolies that impose exploitative pricing and limit functionality \cite{arguelles2021bars, sobol2018connecting}. As Hofinger and Pflegerl note \cite{hofinger2024reality}, such digitalization efforts can function as a ‘Potemkin façade,’ making prisons appear modern without addressing the structural digital marginalization of incarcerated individuals.

In response to these limitations, researchers have started to examine prison technologies and proposed alternative designs grounded in incarcerated people's lived experiences. Researchers have examined the use of VR and generative AI to rehearse post-release scenarios and support family connection \cite{teng2019participatory, martinez2025generative, graf2025virtual,martinez2024engaging}. More broadly, digital literacy has been identified as a critical component of successful reentry across economic, social, and health domains \cite{reisdorf2018digital}. HCI scholars have partnered with community organizations to develop web- and mobile-based tools that support reentry services, job searching, and access to resources \cite{ogbonnaya2019towards, grierson2022design, ogbonnaya2025designing}. In this paper, we build on this prior work and especially build on Verbaan et al.'s call to design technologies that promote the well-being and restoration of selfhood for justice-impacted individuals \cite{verbaan2018potentials}.

\subsection{Punitive Orientations in Parole Technologies}
\rev{Although intended to support reintegration, the U.S. parole system has become increasingly punitive and surveillance-oriented since the late 1970s, shifting away from rehabilitation toward control \cite{wiggins2022parole,hughes2001trends,petersilia1999parole}. Failure to meet parole conditions often leads to reincarceration \cite{phelps2023supervision,grattet2008parole}, and nearly half of parolees do not complete the process \cite{scott2011failure,fish2022constitutional}. In practice, parole frequently returns individuals to prison rather than supporting reentry \cite{wiggins2022parole,hughes2001trends}, while also imposing financial burdens that further hinder reintegration \cite{deitch2022rehabilitation}.}

\rev{Technological developments in parole often mirror these punitive and surveillance-oriented dynamics \cite{kilgore2013progress,al2025effects,laqueur2024algorithmic,itani2024automated,siskou2024so}. For example, prior work has studied computational systems such as electronic monitoring, which extend carceral control beyond prison walls \cite{kilgore2013progress,al2025effects}. More recent efforts emphasize efficiency and administrative optimization, including algorithmic cost estimation, transcript anonymization, and machine learning analyses of parole hearings \cite{laqueur2024algorithmic,itani2024automated,siskou2024so}. Rather than challenging existing institutional priorities, these approaches frequently align with surveillance-oriented dynamics such as compliance, risk management, and operational efficiency.}

\rev{FAccT scholarship can provide a lens for understanding the current design patterns of parole technologies, showing that computational systems in high-stakes institutional contexts can reproduce existing power structures when they prioritize institutional objectives over the needs and perspectives of affected communities \cite{barabas2020studying,eubanks2018automating,benjamin2023race}. In the context of parole, this raises concerns about accountability and epistemic power. Research in this area highlights that accountability mechanisms weaken when impacted communities are excluded from system design \cite{cooper2022accountability,metcalf2021algorithmic}, and that privileging institutional metrics over lived experience can contribute to epistemic injustice \cite{widder2024epistemic,klumbyte2022critical}.} \rev{While prior human-centered and participatory research has begun to engage justice-impacted populations in the design of computational systems \cite{teng2019participatory,martinez2025generative,graf2025virtual,martinez2024engaging, lu2026counter}, technologies specifically focused on supporting individuals through the parole process remain underexplored. This paper builds on these traditions by centering the perspectives of formerly incarcerated individuals to examine how they envision technologies that could better support their parole and reentry.}

\section{Methods}
This research is grounded in the lived experiences and long-term commitments of our team. One co-author was incarcerated in the California state prison system for over 25 years and has undergone two parole board hearings. Our work is guided by the principle of "nothing about us without us, everything with us" \cite{charlton1998nothing}. We view this not as a source of bias, but as a strength that enables deeper insight into the parole process. This commitment shaped our study design, analysis, and dissemination by ensuring that justice-impacted voices were central, rather than peripheral, to the production of knowledge. This perspective is particularly important in the current moment. AI-driven risk assessment, automated monitoring, and data-intensive decision support in parole are rapidly expanding \cite{hartmann2021uncertainty,slobogin2021preventive,dressel2018accuracy}. Yet this expansion has unfolded alongside the continued exclusion of justice-impacted individuals from the design of parole technology. As a result, these technologies are not designed for justice-impacted individuals. Instead, they focus mainly on institutional needs, prioritizing surveillance and control while operating with opaque evaluation methods that lack accountability. This context motivated our research question: What kinds of computational tools would justice-impacted individuals want if they were able to define technology's role in the parole process, rather than having it determined by institutional needs alone?

\subsection{Study Design}
\rev{To address this question, we conducted an IRB-approved, asynchronous mail-in survey with formerly incarcerated individuals who had experienced the parole process. Because recounting parole experiences can be emotionally difficult \cite{western2018homeward,harding2019outside,maruna2011reentry}, we adopted trauma-informed research practices drawn from prior work \cite{alessi2023toward,craig2022wiley,hanley2024ethics, johnson2018involving}. We selected an asynchronous, paper-based format rather than synchronous interviews because real-time interviews can pressure participants to respond quickly and may heighten emotional distress when discussing sensitive experiences \cite{dickson2009researching,pearlman2024interviewing,cook2012email, van2013managing}. A mail-in format, by contrast, allowed participants to respond privately, at their own pace, and to pause, reflect, or skip questions without consequence \cite{amri2021utilizing,saarijarvi2021face,junger1999self}. This design choice was particularly important given that some survey questions asked participants to revisit one of the most consequential administrative processes of their lives, namely the process that could determine whether they regained their freedom or remained incarcerated.}

\rev{To further support participant well-being and reduce barriers to participation, we recruited through reentry organizations. This approach fostered trust and participant agency while making trauma-support resources privately accessible \cite{western2018homeward,harding2019outside,towne2023put}. However, rather than just offering resources only upon request, we included an “Additional Support Resources” section directly into our survey (see Appendix), with contact information for national crisis hotlines and reentry organizations. This made support visible at the moment of reflection. Although all 31 participants received these resources, none requested additional support. Participants were given several weeks to return their responses so that they could complete the survey on their own schedule. The survey itself took approximately 35 minutes to complete. The paper-based format also reduced common participation burdens associated with reentry \cite{visher2004returning,skogseth2025physical}. Justice-impacted individuals often face unstable schedules, transportation barriers, caregiving responsibilities, stigma, and uneven access to digital technologies \cite{western2018homeward,harding2013home,nordberg2021transportation}. Allowing participants to complete the survey at their own pace and without requiring access to digital tools likely facilitated participation \cite{ratislavova2014asynchronous,hargittai2019internet}, which was particularly important given that justice-impacted populations are often difficult to recruit and engage in research \cite{atkinson2001accessing,ellis2023we}. We also covered all mailing and shipping costs to further lower barriers. We did not, however, provide monetary compensation. This decision was made to avoid undue influence that could pressure economically vulnerable individuals to participate or shape their responses in exchange for payment \cite{belmont1979belmont}. Our approach to compensation was developed in consultation with reentry partners, justice-impacted collaborators, and prior work in this area \cite{rodrigues2024participation,smoyer2009compensation}.} 

\rev{Because asynchronous surveys limit opportunities for real-time clarification that interviews might otherwise provide \cite{rezabek2000online}, we designed a carefully structured open-ended instrument grounded in prior research on parole and reentry \cite{petersilia1999parole,travis2003families}. This design was also consistent with our trauma-informed methodology \cite{isobel2021trauma,alessi2023toward}, which prioritized minimizing additional emotional burden on participants. For this reason, we did not conduct routine follow-up interviews or clarification requests. Instead, we limited follow-up communication to participant-initiated requests in order to reduce the possibility of causing further distress.}

{The survey consisted of 10 open-ended questions intended to elicit participants' experiences, challenges, and imaginaries related to parole preparation, as well as their perspectives on the role of technology. The questions were informed by prior work on the parole process and reentry \cite{petersilia1999parole,travis2003families,reisdorf2022digital,western2018homeward,miller2021halfway}. Participants were asked to reflect on: (a) the kinds of assistance they most needed during parole preparation; (b) whether and how they collaborated with family members during preparation; (c) challenges and successes in that collaboration; (d) cultural factors that shaped their parole experiences; (e) their current uses of technology; (f) desired technological affordances; (g) concerns about technology; and (h) future imaginaries for technology in parole, including AI-enhanced systems. To support reflection on AI technologies, we provided examples of widely used tools such as ChatGPT and Gemini.} For questions related to technology and participants’ imaginaries, we draw on prior research on co-designing technologies with underserved communities \cite{imteyaz2026co, do2024designing, martinez2024engaging,martinez2025generative,ogbonnaya2025designing}.
\begin{table}[h]
\centering
\caption{Overview of Participants (N = 31). Data are aggregated and binned to preserve anonymity, given the small number of individuals paroled from the California state prison system.}
\label{tab:participants}
\begin{tabular}{lll}
\hline
\textbf{Category} & \textbf{Range / Group} & \textbf{n (\%)} \\
\hline
Age Range & 39--44 & 5 (16\%) \\
          & 45--49 & 9 (30\%) \\
          & 50--54 & 6 (19\%) \\
          & 55--59 & 5 (16\%) \\
          & 60--64 & 6 (19\%) \\
\hline
Years Incarcerated & 10--14 years & 8 (26\%) \\
                   & 15--19 years & 15 (48\%) \\
                   & 20--24 years & 7 (23\%) \\
                   & 25+ years    & 1 (3\%) \\
\hline
Race/Ethnicity & Hispanic & 22 (72\%) \\
               & Black    & 5 (16\%) \\
               & White    & 4 (12\%) \\
\hline
Parole Hearings (binned) & 1--2 hearings & 18 (58\%) \\
                         & 3 hearings    & 10 (32\%) \\
                         & 4+ hearings   & 3 (10\%) \\
\hline
\end{tabular}
\end{table}
\subsection{Participants}
We recruited participants through community organizations that support reentry and parole preparation. To broaden our reach, we also used snowball sampling, in which initial participants referred others who had recently navigated parole. This combined approach helped us access a population that is often difficult to reach due to barriers of trust, stigma, and institutional gatekeeping \cite{atkinson2001accessing}. Recruitment continued until we reached thematic saturation, at which point enrollment ended. In total, 31 participants completed the survey. Participants ranged in age from 39 to 64, with the largest groups in their mid-40s and mid-50s. They reported a median incarceration length of 17 years (mean $\approx$ 17.0; range: 10--25+), and nearly all had experienced multiple parole hearings (median 2; mean $\approx$ 2.35). Most participants identified as Hispanic, with smaller numbers identifying as Black or White. All participants were male and had been incarcerated within the California Department of Corrections and Rehabilitation (CDCR). Because the number of individuals paroled from California state prisons each year is relatively small, reporting detailed person-level attributes could risk re-identification. To preserve anonymity, we report only binned, aggregate statistics (see Table \ref{tab:participants}).

\subsection{Data Analysis}
\rev{We analyzed participants' responses using qualitative thematic analysis, following established approaches in the literature \cite{braun2006using,braun2019reflecting}. Two researchers independently conducted open coding of all survey responses to identify salient concepts, recurring themes, and notable tensions within the data. We then iteratively refined a shared codebook through discussion and consensus, collapsing overlapping codes and clarifying distinctions between categories. After finalizing the codebook, we re-coded all responses and resolved disagreements through collaborative discussion to ensure consistency and reliability. We also conducted pattern-level analysis to identify recurring themes as well as negative cases and points of tension across participants' accounts \cite{guest2011applied}.}

\rev{This analytic process allowed us to surface both common patterns and unique perspectives that pointed to potential design directions. Throughout the analysis, we remained attentive to the socio-cultural contexts of parole preparation and to how participants connected their lived experiences to broader concerns about technology and justice. Interpretive validity was further strengthened by the research team's positionality. One co-author has lived experience with incarceration and parole and contributed to both the study design and the interpretation of findings, serving as a domain-expert check, consistent with reflexive qualitative research practices \cite{braun2019reflecting}.}

\section{Results}
\rev{Participants described parole as a high-stakes administrative process defined by encounters with institutional power. Securing release did not depend solely on completing required programs or demonstrating personal transformation. It depended on whether that transformation was recognized as credible by decision-makers who hold authority to define what rehabilitation and readiness should look like. Participants consistently emphasized that parole operates through evaluative power: commissioners interpret behavior, language, education, and demeanor through cultural and institutional lenses that determine whether change appears authentic or suspect. Prior scholarship similarly shows that incarcerated individuals learn to anticipate how authorities evaluate them and adapt their behavior accordingly \cite{werth2011envisioning,maruna2011reentry,crewe2009prisoner}. Within this structure, participants described developing deliberate strategies to get ahead of how they would be evaluated. They read hearing transcripts, debriefed peers, profiled commissioners, rehearsed responses, and carefully curated documentation of their achievements. These practices were neither resistance nor passive compliance. Instead, they reflect what we conceptualize as \emph{``strategic agency''}: the disciplined effort to understand how institutional power operates and to build the capacity to move within it toward freedom. In this context, agency is not expressed through opposition but through learning how power reads, measures, and validates change.} 

\rev{Participants' technological imaginaries extended naturally from these navigation strategies. Rather than imagining tools to reduce surveillance or automatically detect bias, participants envisioned technologies that would help them work within the parole system more effectively: tools that clarify board expectations, translate abstract parole concepts into culturally familiar terms, organize progress documentation into board-recognized categories, and support systematic hearing preparation. These imagined technologies function as navigational infrastructure, resources that help justice-impacted individuals and their families anticipate how they will be evaluated and respond to institutional power in ways more likely to be seen as credible.} 

\rev{The themes that follow show how strategic agency develops through participants’ efforts to navigate cultural stereotyping, build informal knowledge networks, and mobilize family support for parole hearings. For each theme, we first provide a description and then present the computational tools participants imagined to address the challenges associated with that theme.}

\subsection{\rev{Theme I: Strategic Agency in Navigating Cultural Stereotyping in Parole Evaluation}}
\subsubsection{\rev{Theme Description}} \rev{Participants described how cultural identities shaped parole decision-making and how they developed strategic agency in response to these dynamics. Cultural bias and misunderstanding were described as active barriers in parole preparation and hearings, influencing what participants felt able to say, how their statements would be interpreted, and whether the parole board would view their rehabilitation and transformation as credible. As one participant stated: \emph{``...being Hispanic worked against us throughout my trial, while incarcerated, and I feel the same way about the board process...''} (P\_4).} \rev{Participants described two ways cultural background affected their parole process:}

\paragraph{1) \rev{CULTURAL STEREOTYPING BY THE PAROLE BOARDS}}
\rev{Parole board commissioners held stereotypes about how people from certain cultural backgrounds should speak and behave. When incarcerated individuals did not match these expectations, they could be seen as inauthentic or dishonest, which in some cases contributed to parole denial. One participant explained that his cultural background made it harder for the parole board to recognize his personal growth and transformation: \emph{``Because we're minority, I feel that being Hispanic, we're judged based on the past and being ex-gang members. A lot of times, we're already seen as threats in the board members eyes, regardless of our change.''} (P\_13). In these accounts, parole board commissioners were perceived as entering hearings with preconceived views tied to cultural background that shaped outcomes more than evidence of rehabilitation.} 

\rev{Participants also emphasized a cultural double bind. On the one hand, efforts to counter stereotypes (for example through education, reflection, or articulate speech) could be discounted. On the other hand, those same efforts could be reframed as manipulation rather than growth. As P\_31 explained, people who can articulate themselves well are sometimes seen as \emph{``superficial, manipulative, or lacking remorse,''} as if education were undertaken to deceive rather than to rehabilitate. Similarly, P\_17 summarized how speech and education could be treated as suspicious: \emph{``The board is such a subjective process that certain commissioners expect us to sound ignorant. When we are educated it's perceived as manipulative, even a threat.''}} \rev{Note that these dynamics can create a double bind. Being highly prepared may be interpreted as manipulative, especially for individuals from certain cultural backgrounds. At the same time, conforming to expected cultural stereotypes can also be used against them, as it may be seen as evidence that they have not changed.}

\rev{In response, participants described developing strategies to anticipate how stereotypes might shape commissioners' interpretations and to adjust their preparation accordingly. For example, P\_29 explained that he was warned his education could \emph{``work against''} him because it could heighten expectations, suggesting that participants attempted to plan for how their transformation would be interpreted under biased cultural frames.}

\paragraph{2) \rev{CULTURAL DISCONNECT WITH PAROLE RELATED CONCEPTS}}

\rev{The parole board also expected incarcerated individuals to understand and engage with specific concepts around rehabilitation and criminal thinking, but these concepts were unfamiliar in some communities or carried different meanings in participants' cultural contexts. As a result, incarcerated individuals struggled to engage with these concepts in the forms expected by the board and were consequently denied parole. For example, P\_15 described how parole expectations relied on concepts that were not taught or commonly used in his community: \emph{``They [parole board] expected me to understand things back then that I understand today, but were not taught in my community or culture...''} Participants described these gaps as making it more difficult to demonstrate board-valued constructs, such as “insight into one’s crime,” in ways that parole commissioners would recognize as credible.} \rev{These cultural dynamics required participants to understand and present their transformation using concepts that aligned with the cultural expectations of the commissioners. As a result, participants described developing strategies to learn these cultural concepts and to communicate their experiences in ways that would be recognized and valued by the parole board.}
\subsubsection{\rev{Theme Imaginaries: Culturally-Aware Technologies for Parole Preparation}}
\rev{Participants’ technological imaginaries reflected these strategies for navigating culturally shaped evaluation. Rather than calling for tools that would primarily audit or expose commissioners’ cultural biases, participants envisioned technologies that would strengthen their strategic agency by helping them navigate a process in which cultural frames shape what counts as credible rehabilitation.} \rev{Participants envisioned tools that help incarcerated individuals recognize the cultural expectations embedded in parole evaluation, anticipate how their speech, demeanor, and narratives may be interpreted through stereotyped lenses, and translate board-valued concepts into culturally familiar and usable terms.} \rev{For this purpose, participants described two main design directions:}

\paragraph{1) \rev{TOOLS FOR TRACKING PRO-SOCIAL ACTIVITIES AND RESISTING CULTURAL BIAS}}

\rev{Participants proposed tools that could track and quantify pro-social activities to produce board-legible evidence, so that accounts of transformation may rely less on culturally contingent impressions and more on structured records. For example, P\_8 described an app that would \emph{``personally monitor and upload all our certificates and progress''} to ensure the parole board could see their full record. Participants framed documentation as a way to reduce reliance on culturally loaded impressions about how a ``rehabilitated'' person should speak or behave by foregrounding concrete evidence of change that aligns with institutional criteria. Several participants extended this idea to AI-enhanced tools that map documented activities into board-recognized categories and standards, helping translate everyday forms of growth into the evaluative language the board uses. P\_26 envisioned an app where an AI would \emph{``put everything into each of the elements that matter for the BPH [Board of Parole Hearings]''} and help indicate when suitability standards were met. Participants suggested that this type of technology could reduce arbitrary denial by making evidence of progress more visible and easier to evaluate through structured records. They also argued that these records could increase accountability. If an AI-enhanced app documents strong evidence of transformation, commissioners who deny parole would need to offer clearer justification for overriding that evidence. As P\_1 noted, if a commissioner went against the AI's assessment, they would have to \emph{``clearly articulate why''}.}

\paragraph{\rev{2) CULTURALLY AWARE TOOLS FOR NAVIGATING THE PAROLE PROCESS}}
\rev{Because incarcerated individuals and commissioners often came from different cultural backgrounds, communication between them could be difficult. Participants therefore envisioned tools that could bridge these gaps by making parole concepts and expectations culturally accessible. Participants described their parole preparation as requiring fluency in institutional language and norms that may not match how accountability, change, and rehabilitation are discussed in their own communities. Rather than scripting answers, participants described interactive systems that could support learning how commissioners evaluate responses and what cultural cues commissioners treat as credible. In this sense, these tools were imagined as something that could scaffold \emph{strategic agency} by helping incarcerated individuals anticipate how their words and stories may be interpreted, and by strengthening their capacity to communicate transformation in forms that are legible within the parole board's power structure. For example, P\_14 proposed a question-and-answer feature that helps individuals \emph{``get a feel for how the commissioners think''} and emphasized that this is \emph{``understanding the game, not faking it''}. Others imagined \emph{``an app to create a study guide that would help inmates articulate ``insight''} (P\_21).} \rev{Participants suggested that these tools could help them learn how the board defines and evaluates key concepts, and then prepare them on how to communicate their own experiences using the terms and narratives that commissioners expect. Overall, participants framed these tools as resources that translate culturally grounded experiences of reflection and transformation into institutional language, helping incarcerated individuals navigate high-stakes evaluations within power structures.}

\subsection{\rev{Theme II: Strategic Agency Through Informal Information Networks}}

\subsubsection{\rev{Theme Description}}\rev{While the first theme focused on developing strategic agency to navigate power structures shaped by cultural dynamics, this theme centers on how incarcerated individuals develop strategies to navigate the parole process through informal information networks. These networks become especially important because official guidance from prisons and parole boards is largely absent. Rather than receiving clear instruction from these institutions, participants described how incarcerated individuals typically learn how parole works through peers, personal hearing transcripts, and shared experiences. As P\_10 explained: \emph{``the only information I received about the board process was obtained from other inmates who had been to the board, I felt ill-informed.''}} 

\rev{Participants described building their own collective knowledge systems to reduce uncertainty in a process where outcomes depend heavily on how commissioners interpret and evaluate them. Through conversations, shared experiences, and document analysis, incarcerated individuals attempted to understand the patterns, expectations, and preferences that shape parole decisions. For example, P\_15 described debriefing peers after their parole hearings to create commissioner profiles, including their \emph{``pet peeves''} and \emph{``line of questions''}, and explained that this process was supported by reading parole transcripts. P\_25 described similar practices, noting that incarcerated individuals \emph{``created profiles on commissioners''} and shared information about reentry resources. By pooling information and documenting patterns in questioning and evaluation, incarcerated individuals attempted to anticipate how they would be assessed and better prepare for encounters with institutional authority. Participants also shared that these knowledge networks were often sustained through family members and friends outside prison, who helped gather information, maintain documents, and connect incarcerated individuals to resources. As a result, individuals without such external support faced significant disadvantages in navigating the parole process. As P\_20 stated, those with \emph{``no one outside to help''} with housing, resources, and networking were \emph{``at a complete disadvantage in the parole process''}.}

\subsubsection{\rev{Theme Imaginaries: Tools to Address Information Gaps in Parole Preparation}}

\rev{Participants envisioned technologies that could extend and stabilize these informal knowledge networks. For example, P\_2 described a desire for multimodal guidance such as board etiquette videos and packet preparation courses, emphasizing that content should be digestible for the broader prison population. Participants also envisioned an \emph{``accessible database that stores all felon friendly employers''} (P\_12), along with reentry resources that could support parole planning. In addition, they proposed an \emph{``online forum to do mock hearings [among incarcerated individual] and share each others' board experience would be a useful app''} (P\_18).} \rev{Participants framed these tools as ways to expand access to the knowledge needed to prepare for parole and better understand how the board evaluates candidates. By making institutional expectations more visible and shareable, these technologies were imagined as supporting strategic agency in parole navigation. The tools could help incarcerated individuals anticipate questioning, prepare their narratives, and engage more effectively with the decision-making authority of commissioners.}

\subsection{\rev{Theme III: Strategic Agency Through Family Support Networks}}

\subsubsection{\rev{Theme Description.}}\rev{The first two themes showed how participants navigated culturally shaped power structures and how they built informal information networks to help them move through those structures. This theme extends that analysis by focusing on the friends and family networks that incarcerated individuals rely on to navigate parole more effectively.} \rev{In particular, participants described how strategic agency was often developed collectively through friends and family who helped manage the administrative demands of parole preparation. They emphasized that preparing for parole is not an individual effort alone, but often a form of collective labor carried out by support networks in the free world. This labor was described as unpaid, essential, and largely unrecognized by the parole system. It included assembling documentation, coordinating with organizations, researching housing and services, and preparing support letters. For example, P\_25 explained, family members researched transitional housing and forwarded emails to support organizations to assist in parole transition.} \rev{Families often also had to learn an unfamiliar system while completing time-intensive administrative tasks under high stakes. In doing so, these support networks developed their own forms of strategic agency as they navigated institutional requirements, timelines, and documentation practices.} \rev{P\_16 explained that the hardest part was helping family members understand \emph{``the complexities that come with the [parole] process''}. Families often felt overwhelmed by the parole process and by the specialized language and norms of hearings. As P\_22 described, the parole hearing process felt \emph{``like speaking another language''}. Participants also described additional risks that shaped whether families could safely participate. For example, P\_9 explained that his family \emph{``were just as scared to help, just because of their nationality''}. Similarly, P\_26 described language barriers: \emph{``The fact that most of my family is predominantly Spanish speaking [meant that] most of the board process was lost on them in translation. It also was difficult for them to communicate with the necessary support organizations...''}}

\subsubsection{\rev{Theme Imaginaries: Family-Centered Tools for Parole Preparation}}

\rev{Participants envisioned technologies that help families navigate the parole system and support incarcerated individuals more effectively. For example, P\_31 proposed an app that breaks down \emph{``complex or abstract concepts''} into lay terms and provides examples of the documents needed for board preparation. Participants also suggested tools that help families learn \emph{``where they could be helpful in the parole hearing process''} (P\_11).} \rev{Multilingual support, particularly Spanish, was repeatedly emphasized. Participants wanted translated definitions and resources, including \emph{``definitions of parole terms provided, sample support letters, forms, transitional housing information, and other reentry resources,''} (P\_17).} \rev{Some also wanted bilingual document generation so materials could \emph{``be better understood by those in my family who are more comfortable with Spanish...,''} (P\_23).} \rev{Finally, participants envisioned official tools from state prisons \emph{``that promotes education, networking and family reunification through technology,''} (P\_6), as well as technologies that help incarcerated individuals \emph{``to familiarize themselves with an ever changing society and familiarize them with modes of communication''} (P\_30).} \rev{Participants framed these tools as supporting reentry by strengthening families' ability to participate in parole preparation and by helping formerly incarcerated individuals reconnect with evolving communication norms and community life after release.}
\section{{Discussion}}
In this paper, we examined what computational tools justice-impacted individuals would want if they could define technology's role in the parole process themselves. For this purpose, we conducted a survey with 31 justice-impacted individuals about their parole experiences, focusing on the key pain points they encountered. Based on their experiences, participants envisioned computational tools that could help currently incarcerated individuals navigate the parole process more effectively and limit the challenges they faced.

\subsection{\rev{Strategic Agency for Navigating Power Structures}}
\rev{Participants described how parole boards evaluate credibility and rehabilitation through cultural norms that shape how incarcerated individuals are expected to speak, behave, and demonstrate remorse. We observed that these expectations created a double bind. When individuals conform to commissioners’ expectations, they may be seen as fitting a stereotypical “criminal type” and therefore as lacking genuine rehabilitation. However, when they exceed those expectations by demonstrating education, insight, or articulate communication, those same behaviors may be reinterpreted as manipulation or impression management rather than authentic growth. For example, participants described stereotypes that frame Latino incarcerated individuals as uneducated or as speaking broken English. When a Latino person appears at a hearing with educational achievements and communicates in a polished manner, commissioners may read this not as evidence of transformation but as strategic performance. In this way, stereotyping does not simply obscure change; it actively reshapes the criteria of credibility, leaving justice-impacted individuals without a reliable way to have their rehabilitation recognized.} \rev{Prior work shows that parole boards often rely on subjective, potentially biased “gut feelings” rather than consistent, evidence-based evaluation \cite{ruhland2020philosophies,caplan2007factors}. Our findings advance this literature in two important ways. First, the double bind described by participants shows that bias does not only punish stereotype-consistent behavior; it can also penalize stereotype-disconfirming performances of growth. This extends existing accounts of parole bias by showing that even efforts to exceed institutional expectations may be read against incarcerated individuals. Second, our findings highlight how cultural frames can render transformation illegible to decision-makers. This might help explain a recurring puzzle in parole scholarship \cite{shinnar1975effects,jones2007probation}: why some individuals are repeatedly denied parole despite completing required programs, maintaining strong institutional records, and meeting formal milestones \cite{caplan2007factors,kokkalera2024not,gottfredson1979parole}. If change is not recognized as authentic within a biased credibility framework, then institutional achievement alone may be insufficient to overcome stereotyped judgments about rehabilitation.}

\rev{Our findings also contribute to broader carceral scholarship on institutional power and self presentation \cite{werth2011envisioning,maruna2011reentry,crewe2009prisoner}. Werth shows that parole institutions construct individuals as responsibilized subjects who are expected to demonstrate transformation while still being treated as inherently risky \cite{werth2011envisioning,werth2012m}. This creates a structural tension in how people must present themselves to authorities. Maruna similarly argues that successful self-presentation in parole often depends on constructing coherent redemption narratives that align with institutional expectations of what “real” change looks like, even when those expectations do not fully reflect lived experience \cite{maruna2011reentry}. Crewe further shows that incarcerated individuals develop a form of “soft power” awareness, becoming highly attuned to how authorities interpret their self-presentation during parole hearings  \cite{crewe2009prisoner}. Our findings extend this prior work by showing that self-presentation for parole is not only about understanding institutional expectations, but also about recognizing how those expectations are filtered through racialized and cultural stereotypes that make credibility unstable. Participants described having to show transformation while also anticipating how that transformation would be read through stereotyped cultural lenses. They therefore had to adapt their self-presentation to credibility standards that were powerful, consequential, and often never explicitly stated. Because participants saw navigating these dynamics as both difficult and high stakes, they imagined computational tools that could help incarcerated individuals better understand and move through these power structures. Taken together, our findings suggest that parole is not simply about self-presentation. It is also about exercising \emph{strategic agency}: developing the capacity to navigate institutional power, anticipate its judgments, and present oneself in ways that can be recognized as credible within culturally shaped systems of evaluation.}

\rev{To further interpret this form of agency, we draw on the work of sociologist and critical theorist Mahmood \cite{mahmood2016feminist}. Mahmood argues that agency is not limited to resisting power, but can also emerge through practices that appear docile or subordinate on the surface. She illustrates this through the example of a pianist who achieves mastery not through unstructured self-expression or opposition to authority, but through discipline, receptivity to instruction, and sustained practice. This docility is not passive surrender; it is an active, effortful process through which new capacities are developed. Our participants described an analogous form of preparation: reading hearing transcripts, analyzing commissioners’ questioning styles, rehearsing responses, and learning to present personal transformation in culturally legible ways. Our contribution here is to connect how these practices can be understood not as mere compliance, but as the development of practical capacities for navigating a high-stakes institutional process. These practices required substantial effort, and participants viewed them as essential to improving their chances of securing release.} \rev{The technological imaginaries of our participants reflected this same orientation. Rather than asking for tools that automate decisions, detect bias after the fact, or simply intensify surveillance, participants envisioned computational tools that scaffold learning, translation, and preparation. They imagined computational systems that could help incarcerated individuals understand how parole boards evaluate them, translate parole concepts into culturally accessible forms, document transformation in institutionally legible ways, and support families in knowing when and how to provide assistance. Our novel contribution is to connect that the technologies participants imagined are oriented toward supporting the development of \emph{strategic agency}. We therefore call for parole technologies that move beyond the dominant design logics of surveillance, compliance, or collective action alone \cite{ziosi2024evidence,partnership2019report,10.1145/3491101.3503820,armstrong2020you,savage2016botivist}, and instead support incarcerated individuals in building the capacities needed to navigate power structures in pursuit of freedom.}

\subsection{\rev{Strategic Agency as a Contribution to FAccT Discussions around Power}}
\rev{Our findings connect to and extend prior FAccT conversations about power and computational systems. Like many systems examined in FAccT, parole is deeply shaped by institutional power \cite{widder2023s,ehsan2022algorithmic,metcalf2021algorithmic,kim2023organizational}. Parole board commissioners and prison staff control how parole and related technologies are designed and enacted. Similar to software engineers who may recognize unfairness but lack the institutional power to refuse building such systems \cite{widder2023s}, incarcerated individuals often lack formal means to challenge unfairness in parole, limiting accountability. This dynamic mirrors broader FAccT findings \cite{ramesh2022platform,johnson2024fall,cooper2022accountability}, where those most affected by computational systems frequently have limited power to contest harmful or unfair outcomes.}

\rev{Our findings also resonate with prior FAccT work on power and harm \cite{boag2022tech,devrio2024building,10.1145/3531146.3533194}, while extending it in important ways. Consistent with this literature, we find that incarcerated individuals are not passive recipients of harm; instead, they develop responses under highly constrained conditions. However, the form of agency we observe differs from those emphasized in prior work. For example, DeVrio et al. \cite{devrio2024building} highlight responses such as building alternatives, shifting power relations, or contesting harm from below, which assume some capacity to challenge or reconfigure governing systems. In contrast, our participants more often described strategies oriented toward navigating institutional power rather than directly contesting it. This difference likely reflects the parole context, where contestation can be risky, infeasible, or even detrimental to one’s chances of release. Accordingly, participants did not describe filing complaints or organizing collective action through technology. Instead, they exhibited \textit{strategic agency}: a form of response focused on anticipating institutional judgments and learning how to navigate them effectively. We propose \textit{strategic agency} as a distinct response to harm that becomes especially salient under conditions of extreme power asymmetry, high stakes, and limited avenues for contestation. In such contexts, individuals may not seek to transform the system, but rather to develop the capacities needed to navigate it in ways that improve their chances of success. In this way, our findings extend prior FAccT taxonomies of responses to algorithmic harm \cite{devrio2024building}.}

\rev{Finally, our work builds on FAccT scholarship that critiques computational systems used on justice-impacted individuals through the lenses of fairness, accountability, and transparency, especially work on bias, auditability, and the harms of algorithmic decision-making \cite{ehsan2022algorithmic,selbst2019fairness,raji2020closing}. This literature has been essential in showing how technology can reproduce institutional inequality and operate as a tool of control \cite{benjamin2023race,Zilka23}. However, it has paid less attention to how justice-impacted individuals themselves envision computational tools that could support them. Our work addresses this gap by examining how formerly incarcerated individuals imagine such technologies in the context of parole. In the next section, we present design recommendations grounded in participants’ insights, with a particular focus on systems that support strategic agency and navigation of power.}

\subsection{{\rev{Design Implications.}}} 
\rev{One major design implication of our findings is that our participants wanted computational tools that help incarcerated individuals build \emph{strategic agency} to navigate the power structures shaping parole. Participants were not asking for systems that automate decisions or reduce effort. Instead, they envisioned tools that help them understand how parole boards evaluate cases, identify what forms of evidence matter, prepare for hearings, and present rehabilitation in ways more likely to be recognized as credible. In this sense, we argue that parole-focused computational systems should equip incarcerated individuals with the knowledge, interpretive skills, and communication capacities needed to navigate high-stakes institutional evaluation. These systems should also translate institutional language and cultural expectations by explaining parole concepts in accessible terms and clarifying how commissioners interpret speech, demeanor, and narratives of change.} \rev{Our findings further suggest that such technologies should be grounded in the lived experiences of justice-impacted individuals. Participants revealed forms of knowledge that are rarely captured in official policies or technical specifications \cite{metchik1988parole,proctor1999new}, including informal expectations, cultural cues, and credibility judgments that shape decision-making in practice. Designing from these lived experiences would position justice-impacted knowledge as a legitimate basis for identifying harms and designing interventions.} 

\rev{At the same time, deploying these types of computational systems within prison infrastructure may be difficult \cite{jewkes2016brave,reisdorf2016b,zivanai2022digital,mckay2022carceral}. Authorities may view computational tools that help individuals analyze evaluation criteria or navigate institutional expectations as enabling manipulation rather than rehabilitation \cite{werth2012m,caplan1992parole,ruhland2020philosophies}. Even if permitted, such systems could be co-opted to reinforce institutional norms \cite{wang2025legality,burchell2013csr}. For example, if an AI tool helps individuals “speak the board’s language,” that style could become a new expectation by the board, reducing the tool’s intended benefit.} 

\rev{Given these constraints, we argue that these technologies may be more viable as “alternative social platforms” \cite{10.1145/2940299.2940314,gehl2015case,gow2022turning}, meaning systems that operate alongside existing infrastructures without requiring institutional approval \cite{grevet2015piggyback,10.1145/3686899,10.1145/3476060}. One promising approach for enabling these types of technologies could be through the use of web plugins \cite{picazo2020after,cao2020activist,firmenich2022engineering}, which have been shown to support gig workers without requiring official platform approval and without being easily appropriated by those in power \cite{10.1145/3476060,10.1145/3686899,savage2020becoming,chiang2018crowd, saito2019turkscanner}. This suggests a useful model for parole tools.} \rev{However, because incarcerated individuals typically cannot install unofficial or alternative software \cite{reisdorf2016b,jewkes2016brave,mckay2022carceral}, a more practical approach is to design these tools for their friends and family members. Our results and prior work show that families already play a central role in parole preparation \cite{visher2004returning,travis2003families,western2018homeward}, and plugins could provide support within the tools they already use \cite{grevet2015piggyback,firmenich2022engineering}. These tools could help families build strategic capacity by explaining how parole boards evaluate rehabilitation, what evidence matters, how institutional language should be interpreted, and how to support credible presentations of change.} \rev{Family members could then share this knowledge through official prison communication channels such as in-person visits, letters, phone calls, or tablet messaging systems \cite{folk2019behind,wang2021research}. We also envision social tools similar to Ameelio that support preparing and mailing educational materials to incarcerated individuals \cite{annis2022connecting,roose2021good}.} \rev{Beyond information sharing, these types of social tools could also strengthen family connections and make their often invisible labor more visible, for example through badges or other recognition features \cite{comfort2019doing,smith2025bridging}.} \rev{For incarcerated individuals without strong outside support networks, similar computational tools could be used by volunteers in reentry organizations. These tools could scaffold parole preparation while helping volunteers build skills for directly working with incarcerated individuals \cite{elisha2025civil,kjellstrand2023importance}. These support models could also extend to prison pen-pal networks \cite{latteri2025writing,mejiaodonnell2019exploring,mutis1999diary}. Based on our findings, these tools should also be multilingual and support the languages preferred by friends and family members. Fig.~\ref{fig:teaser} provides an overview of our paper's proposed design direction.}\\

{{\bf Limitations.}} \rev{Our study has limitations. While the asynchronous survey format reduced emotional pressure and allowed participants to respond at their own pace, it limited opportunities for real-time clarification. We addressed this through a carefully designed open-ended instrument grounded in parole and reentry literature \cite{petersilia1999parole,travis2003families,reisdorf2022digital,western2018homeward,miller2021halfway}, as well as through interpretation informed by a co-author with lived experience of incarceration and parole.} \rev{Additionally, all participants were male, formerly incarcerated in California, and most identified as Hispanic, which may limit transferability across gender, racial groups, and jurisdictions. Our recruitment strategy, which relied on community organizations and snowball sampling, may also have favored individuals with stronger support networks. However, our contribution is primarily theoretical rather than statistically generalizable. Given the challenges of recruiting and engaging justice-impacted individuals in research \cite{kitt2020barriers,randolph2025integration}, we view these findings as an important first step. We present strategic agency not as a universal claim, but as an analytic lens for understanding how underserved individuals engage with computational tools in high-stakes bureaucratic settings.} \rev{Although our empirical context is parole, the broader dynamics we identify are not unique to this setting. Similar dynamics can usually be found in other high-stakes bureaucratic systems where outcomes depend on discretionary evaluation and credibility judgments, such as immigration hearings, welfare eligibility, or asylum decisions \cite{eagly2014remote,ryo2017legal,sainsbury2012welfare,fassin2013enforcing}. In this sense, our study offers a first step toward understanding how computational systems may serve as cultural bridges in contexts where credibility shapes access to freedom.}

\section{Conclusion}
\rev{This paper examines how justice-impacted individuals envision computational tools for navigating parole, a high-stakes process shaped by concentrated authority and limited opportunities for contestation. Rather than calling for tools that automate decisions or simply expose bias, participants envisioned technologies that help them understand how parole boards evaluate credibility, interpret institutional expectations, and prepare to present their transformation in institutionally legible ways. We conceptualize these practices as \emph{strategic agency}: the development of capacities for navigating institutional power in pursuit of freedom. Our work contributes to FAccT scholarship by showing how parole operates as a socio-technical process shaped by institutional power, by identifying strategic agency as a key response under extreme power asymmetries, and by showing why the lived experiences of justice-impacted individuals are essential for designing computational tools that can support empowerment and freedom.}\\

{\bf Acknowledgments.} Special thanks to the justice-impacted individuals who participated in the study and the anonymous reviewers. This work was partially supported by NSF grants 2339443 and 2403252.

\bibliographystyle{ACM-Reference-Format}
\bibliography{references, lit1}
\newpage
\newpage

\section{\rev{APPENDIX: Survey Instrument: Parole Preparation, Support, and Technology}}

\rev{This appendix presents the full survey instrument used in our study. The survey consisted of demographic questions followed by open-ended questions.}

\subsection{\rev{Demographic Questions}}

\textbf{\rev{1. What is your age range?}}
\begin{itemize}
    \item $\square$ \rev{18--24}
    \item $\square$ \rev{25--31}
    \item $\square$ \rev{32--38}
    \item $\square$ \rev{39--44}
    \item $\square$ \rev{45--49}
    \item $\square$ \rev{50--54}
    \item $\square$ \rev{55--59}
    \item $\square$ \rev{60--64}
    \item $\square$ \rev{65 or older}
    \item $\square$ \rev{Prefer not to answer}
\end{itemize}

\textbf{\rev{2. How do you describe your gender identity?}}
\begin{itemize}
    \item $\square$ \rev{Man}
    \item $\square$ \rev{Woman}
    \item $\square$\rev{ Non-binary / Genderqueer}
    \item $\square$ \rev{An identity not listed here:} \underline{\hspace{5cm}}
    \item $\square$ \rev{Prefer not to answer}
\end{itemize}

\textbf{\rev{3. How do you describe your race or ethnicity? (Select all that apply.)}}
\begin{itemize}
    \item $\square$ \rev{American Indian or Alaska Native}
    \item $\square$ \rev{Asian}
    \item $\square$ \rev{Black or African American}
    \item $\square$ \rev{Hispanic or Latino}
    \item $\square$ \rev{Native Hawaiian or Other Pacific Islander}
    \item $\square$ \rev{White}
    \item $\square$ \rev{Another identity:} \underline{\hspace{5cm}}
    \item $\square$ \rev{Prefer not to answer}
\end{itemize}

\textbf{\rev{4. Since turning 18, approximately how many total years were you incarcerated?}}
\underline{\hspace{5cm}}
\textbf{\rev{5. How many parole hearings did you go through?}}
\begin{itemize}
    \item $\square$ \rev{1 hearing}
    \item $\square$ \rev{2 hearings}
    \item $\square$ \rev{3 hearings}
    \item $\square$ \rev{4 or more hearings}
    \item $\square$ \rev{Prefer not to answer}
\end{itemize}

\subsection{\rev{Open-Ended Questions}}

\textbf{\rev{1. Parole Hearing Experience}}

\textbf{a.} \rev{Please describe your experience during your parole hearing(s). Particularly, what stood out to you about how the board evaluated you?}

\vspace{0.5cm}

\textbf{b.} \rev{How did you decide what information to share or how to present yourself to the board?}

\vspace{0.5cm}

\textbf{\rev{2. Needed Support}}

\rev{What kinds of assistance did you most need while preparing for parole?}

\vspace{0.5cm}

\textbf{\rev{3. Changes Across Hearings}}

\rev{Did you go through more than one parole hearing?}

\textbf{a.} \rev{If so, what changed in how you prepared between hearings?}

\vspace{0.3cm}

\textbf{b.} \rev{What were you hoping the board would see or understand about you differently during later hearings?}

\vspace{0.3cm}

\textbf{c.} \rev{Besides official instructions, how did you learn what the board wanted to see?}

\vspace{0.3cm}

\textbf{d.} \rev{How did that information change your approach for the hearing?}

\vspace{0.5cm}

\textbf{\rev{4. Family Support}}

\rev{Did any family members help you prepare for parole? If so, how did they help?}

\textbf{a.} \rev{What worked well when you and your family members prepared for parole together?}

\vspace{0.3cm}

\textbf{b.} \rev{What was most challenging?}

\vspace{0.5cm}

\textbf{\rev{5. Cultural Factors}}

\rev{Were there any cultural factors that shaped your parole experience?}

\textbf{a.} \rev{If so, please describe them.}

\vspace{0.5cm}

\textbf{\rev{6. Technology Use}}

\rev{Did you use any technology during parole preparation or reentry?}

\textbf{a.} \rev{If so, please describe what you used and how it helped or did not help.}

\vspace{0.5cm}
\newpage
\textbf{\rev{7. Desired Technology}}

\rev{What kinds of technologies or features would have been most helpful to support you during parole preparation?}

\vspace{0.5cm}

\textbf{\rev{8. Concerns About Technology}}

\rev{What concerns, if any, would you have about using technology to support people preparing for parole?}

\vspace{0.5cm}

\textbf{\rev{9. Technology and Lived Experience}}

\rev{How could technology better support people preparing for parole, considering their life experiences, including trauma?}

\vspace{0.5cm}

\textbf{\rev{10. AI and Future Tools}}

\rev{Looking ahead, imagine that advanced AI systems (computers that can understand language, answer complex questions, and help organize information) were available to support parole preparation. Examples of current tools like this include ChatGPT or Gemini, which can help people write letters, summarize long documents, or practice for interviews.}

\textbf{a.} \rev{In an ideal world, how would you want these tools to help someone preparing for their parole hearing?}

\vspace{0.3cm}

\textbf{b.} \rev{Based on your experiences, what are specific things these tools should \textbf{not} be allowed to do?}

\subsection{\rev{Additional Support Resources}}

\rev{We recognize that reflecting on the parole process and experiences of incarceration can be emotionally difficult. If you need support, you may consider reaching out to your re-entry center or these national resources:}

\begin{itemize}
    \item \textbf{\rev{National Reentry Resource Center (NRRC)}} \\
    \rev{Phone: (877) 332-1719} \\
    \rev{Website: \url{https://nationalreentryresourcecenter.org}}

    \item \textbf{\rev{Crisis Text Line}} \\
    \rev{Text HOME to 741741 (24/7 support)}

    \item \textbf{\rev{988 Suicide \& Crisis Lifeline}} \\
    \rev{Call or text 988 (24/7 support)}
\end{itemize}

\end{document}